\documentclass[aps, prd, twocolumn, notitlepage, nofootinbib]{revtex4-2}
\usepackage{graphicx}
\usepackage{float}
\usepackage{amsmath}
\usepackage{color}
\usepackage{tensor}
\usepackage{epstopdf}
\usepackage{xcolor}
\usepackage{wasysym}
\usepackage{hyperref}
\newcommand{\Eprint}{\href}
\hypersetup{
    colorlinks,
    linkcolor={blue!50!black},
    citecolor={red!50!black},
    urlcolor={blue!80!black}
}

\usepackage{multibib}

\newcommand{\be}{\begin{equation}}
\newcommand{\ee}{\end{equation}}
\newcommand{\ba}{\begin{eqnarray}}
\newcommand{\ea}{\end{eqnarray}}

\newcommand{\beq}{\begin{equation}}
\newcommand{\eeq}{\end{equation}}
\newcommand{\beqa}{\begin{eqnarray}}
\newcommand{\eeqa}{\end{eqnarray}}
\newcommand{\nn}{\nonumber}
\graphicspath{{plots/}}

\begin{document}

\title{Distinguishing a Slowly Accelerating Black Hole \\by Differential Time Delays of Images}

\author{Amjad Ashoorioon}
\email{amjad@ipm.ir}
\affiliation{School of Physics, Institute for Research in Fundamental Sciences (IPM), P.O. Box 19395-5531, Tehran, Iran}

\author{Mohammad Bagher Jahani Poshteh}
\email{jahani@ipm.ir}
\affiliation{School of Physics, Institute for Research in Fundamental Sciences (IPM), P.O. Box 19395-5531, Tehran, Iran}

\author{Robert B. Mann}
\email{rbmann@uwaterloo.ca}
\affiliation{Department of Physics and Astronomy, University of Waterloo, Waterloo,
Ontario, N2L 3G1, Canada}
\affiliation{Perimeter Institute for Theoretical Physics, Waterloo, Ontario, N2L
2Y5, Canada}

%


\begin{abstract}
Accelerating supermassive black holes, connected to cosmic strings, could contribute to structure formation and get captured by galaxies if their velocities are small. This would mean that the acceleration of these black holes is small too. Such a slow acceleration has no significant effect on the shadow of such supermassive black holes. We also show that, for slowly accelerating black holes, the angular position of images in the gravitational lensing effects do not change significantly. We propose a method to observe the acceleration of these black holes through the gravitational lensing. The method is based on the observation that differential time delays associated with the images are substantially different with respect to the case of non-accelerating black holes.
\end{abstract}

\maketitle

Over the past five years observations from LIGO/VIRGO has provided us with qualitatively new information about our universe via gravitational waves. Black holes have so far been the primary source of gravitational waves, but it is reasonable to expect that future detectors will discover new sources, some of which
will yield information about the early universe.  Cosmic strings -- line-like topological defects that can emerge from some gauge theories during first order phase transitions~\cite{Kibble:1976sj,vilenkin1985} or at the end of brane inflation~\cite{Sarangi:2002yt} -- are one such example.  Such objects can break or fray to produce pair of accelerating black holes~\cite{hr,eardley1995}, which can also be produced in a background magnetic field~\cite{garfinkle1991,hawking1995,dowker1994} or in de Sitter space~\cite{mellor1989,mann1995,dias2004:2,Ashoorioon:2014ipa,ashoorioon2021}. Alternatively, a network of cosmic strings could capture  primordial black holes produced during their formation~\cite{vilenkin2018}. Such black holes would be accelerating due to the tension of the cosmic string.

It is therefore of interest to examine how we might detect accelerating black holes.  Although accelerating supermassive black holes connected to cosmic strings could reside in the centers of galaxies~\cite{vilenkin2018,morris2017}, their velocities must be small ($\lesssim 100\, {\rm km}/{\rm s}$), so that they can contribute to structure formation~\cite{vilenkin2018}. Measurement of their acceleration is therefore a formidable challenge, since this quantity must consequently be very small.

Here we propose a method for detection of black hole acceleration via gravitational lensing, in which rays of light  coming from a source behind the black hole are deflected near it and turn toward an observer. The observer  does not see the  actual location of the source, but rather sees images of it apparently located elsewhere.   Our method exploits the fact that differential time delays associated with lensed images of accelerating black holes substantially differ with respect to their non-accelerating counterparts.

\begin{figure}[htp]
	\centering
	\includegraphics[width=0.5\textwidth]{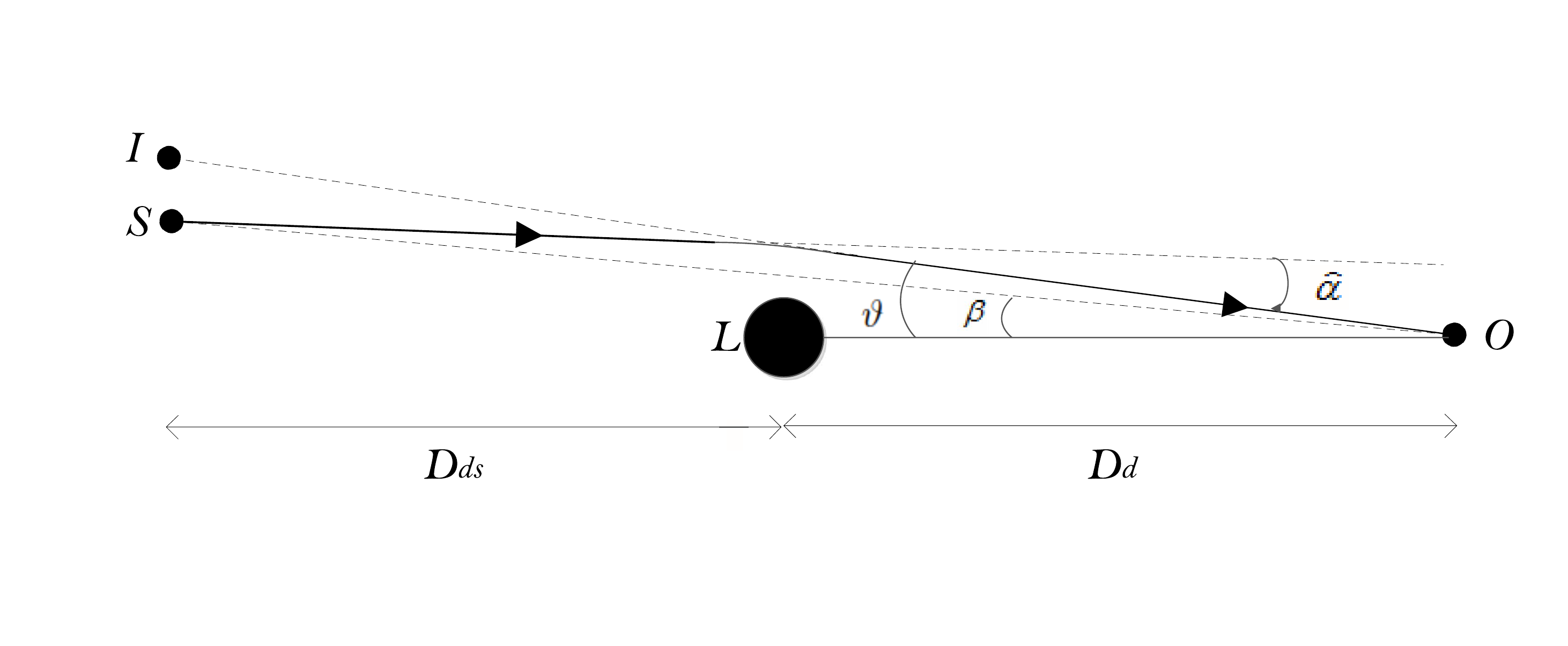}
	\caption{{\it Deflection of light by the black hole}: $S$, $I$, $O$, and $L$, respectively stand for the source, image, observer, and the lens (black hole). The black hole is accelerating in a direction perpendicular to the plane of the figure and behind it there is an acceleration horizon at $r=1/\alpha$, where $\alpha$ is the acceleration, whose sign is not important in our approximation. The path of the light ray remains on the equatorial plane of the black  hole within the small acceleration approximation $\alpha D_s \ll 1$. 
	 $\beta$ is the actual angular position of the source (with respect to the line of sight to the black hole). The black hole  bends the light ray by an angle $\hat{\alpha}$ so that  the observer sees the image at angular position $\vartheta$. $D_d$ and $D_{ds}$ are the distance from lens to observer and from lens to the source, respectively.}
	\label{fig:lens_diag}
\end{figure}

Spacetime around uniformly accelerated black holes can be described by
\begin{eqnarray}
	ds^2&=&\frac{1}{(1+\alpha r\cos\theta)^2} \nn\\
	&\times&\left[-Q(r)dt^2+\frac{dr^2}{Q(r)}+\frac{r^2d\theta^2}{P(\theta)}+P(\theta)r^2\sin^2\theta d\phi^2\right], \label{eqn:metric:griffiths}\nn\\
\end{eqnarray}
known as the C-metric~\cite{kinnersley1970,griffiths2006},
where
\begin{eqnarray}
	Q(r)&=&(1-\alpha^2 r^2)\left(1-\frac{2m}{r}\right), \nn\\
	P(\theta)&=&1+2\alpha m \cos\theta. \label{eqn:tran:griffiths}
\end{eqnarray}
$m$ is the mass parameter and $\alpha$ is interpreted as the acceleration. 
We take the acceleration to be sufficiently small so that a ray of light passing the black hole to the Earth, lies  on the equatorial plane $\theta=\pi/2$ of the black hole during its  evolution. On this plane, the line element \eqref{eqn:metric:griffiths} reduces to
\footnote{ Within the small acceleration approximation, we can first find the geodesic equations and then substitute $\theta = \pi/2$ --- the results will be the same as presented here.}
\be
ds^2=-Qdt^2+\frac{dr^2}{Q}+r^2d\phi^2, \label{eqn:metric}
\ee
where $Q(r)$ is given by \eqref{eqn:tran:griffiths}.  In Fig.~\ref{fig:lens_diag} we   schematically illustrate the setup. Light coming from the source   passes the black hole and is deflected by  an angle $\hat{\alpha}$.  The observer sees the image of the object at angular position $\vartheta$.

The equations governing the geodesics can be obtained using the Lagrangian
\be
\mathcal{L}=\frac{1}{2}g_{\mu\nu}\dot{x}^\mu\dot{x}^\nu=\frac{1}{2}\left(-Q\dot{t}^2+\frac{\dot{r}^2}{Q}+r^2\dot{\phi}^2\right), \label{eqn:lag}
\ee
where the dot denotes differentiation with respect to some affine parameter along the geodesic.
The constants of motion are
$E=-\frac{\partial \mathcal{L}}{\partial \dot{t}}=Q\dot{t}$, and $L_z=-\frac{\partial \mathcal{L}}{\partial \dot{\phi}}=-r^2\dot{\phi}$.

For the null geodesics  $\mathcal{L}=0$, and at the point of closest approach to the black hole, $r=b$,   $\frac{dr}{d\phi}=0$. Using these facts the deflection angle is\footnote{As can be seen from Eq.~\eqref{eqn:tran:griffiths}, the metric function $Q$ changes sign at $r=1/\alpha$. Because of the $1/r$ factor in Eq.~\eqref{eqn:alpha_hat}, the contribution of large $r$ from $\alpha^{-1}$ to infinity is very small in the integral of \eqref{eqn:alpha_hat}--- about one part in one million parts. Therefore we can safely integrate to infinity.}~\cite{weinberg1972}
\be\label{eqn:alpha_hat}
\hat{\alpha}(b)=2\int_{b}^{\infty}\frac{dr}{r\sqrt{\left(\frac{r}{b}\right)^2Q_b-Q}}-\pi,
\ee
where we safely assume that~$1\gg \alpha D_d, D_{ds}\gg b$, where
  $D_d$ and $D_{ds}$ are
the respective distances from observer and source to the black hole.  For simplicity we  take the direction of the acceleration to be perpendicular to the plane of the lens diagram, i.e.~Fig.~\ref{fig:lens_diag}. If the acceleration has a component parallel to this plane, then the third component of the angular momentum is not a constant of motion and the geodesic equations cannot be analytically integrated.

Using the Lagrangian \eqref{eqn:lag} and the fact that $\frac{dr}{dt}=0$ at $r=b$ we find
\be\label{eqn:time_delay}
\tau(b)=\left[\int_{b}^{r_s}dr+\int_{b}^{D_d}dr\right]\frac{1}{Q\sqrt{1-\left(\frac{b}{r}\right)^2\frac{Q}{Q_b}}}-D_s\sec\beta,
\ee
for the difference between the time it takes for  light to travel the physical path from  source to  observer with and without a black hole present, where $\beta$ is the angular position of the source, $D_s=D_d+D_{ds}$ is the distance from observer to the source and $r_s=\sqrt{D_{ds}^2+D_s^2\tan^2\beta}$.

The  Virbhadra-Ellis lens equation~\cite{virbhadra2000}
\be\label{eqn:lens_eq}
\tan\beta=\tan\vartheta-\mathcal{D}\left[\tan\vartheta+\tan(\hat{\alpha}-\vartheta)\right],
\ee
where $\mathcal{D}=D_{ds}/D_s$, yields the image angular position, $\vartheta$, and
\be\label{eqn:impact}
J=\frac{b}{\sqrt{Q_b}}=D_d\sin\vartheta
\qquad
\mu=\left(\frac{\sin\beta}{\sin\vartheta}\frac{d\beta}{d\vartheta}\right)^{-1}.
\ee
are respectively the  impact parameter and image magnification~\cite{virbhadra1998}.

Differentiating Eq. \eqref{eqn:lens_eq} with respect to $\vartheta$ we find
\be
\sec^2\beta\frac{d\beta}{d\vartheta}=\sec^2\vartheta-\mathcal{D}\left[\sec^2\vartheta+\sec^2(\hat{\alpha}-\vartheta)\left(\frac{d\hat{\alpha}}{d\vartheta}-1\right)\right].
\ee
where $\frac{d\hat{\alpha}}{d\vartheta}=\frac{d\hat{\alpha}}{db}\frac{db}{d\vartheta}$. The factor $\frac{db}{d\vartheta}$ can be found by from~\eqref{eqn:impact}, whereas we use~\cite{poshteh2019}
\be\label{eqn:alpha_r}
\frac{d\hat{\alpha}(b)}{db}=-2\int_{b}^{\infty}\frac{1}{\sqrt{\mathcal{F}}}\frac{\partial}{\partial r}\left(\frac{1}{r}\frac{\partial\mathcal{F}}{\partial b}\frac{\partial r}{\partial\mathcal{F}}\right)dr,
\ee
where $\mathcal{F}=\left(\frac{r}{b}\right)^2Q_b-Q$, to compute   $\frac{d\hat{\alpha}}{db}$.

We take the lensing black hole to be of the same mass and distance as M87*. We use numerical methods~\cite{virbhadra2000,poshteh2019} to investigate its gravitational lensing assuming  this black hole is accelerating and compare  image positions, magnifications and (differential) time delays to the case where it is non-accelerating. The mass and distance of M87* have been obtained by the  Event Horizon Telescope Collaboration as $M_{{\rm M87^*}}=9.6\times 10^{12} \, {\rm m}\equiv 6.5\times 10^9 M_{\odot}$ and $D_d=5.2\times 10^{23} \, {\rm m}$~\cite{Akiyama:2019eap}. Whenever we consider an accelerating black hole, we take the acceleration to be $\alpha=10^{-25} {\rm m}^{-1}$, roughly the upper bound obtained by \cite{vilenkin2018} from the development of a cosmic string network with black holes of such  masses as beads in the network. For simplicity we assume that the black hole is halfway between the source and the observer; therefore ${\cal D}=0.5$.  Small changes (of order 1\%) in ${\cal D}$ result in a 0.5\% change in image positions and
a 3\% change in the time delay.

The closer the light ray passes by the black hole, the greater the deflection angle. The small value of the acceleration does not change the value of the deflection angle significantly. In fact, for a fixed value of the impact parameter, the deflection angle of a non-accelerating black hole is greater than that of its accelerating counterpart  by only about 1 part in $10^{13}$. Since the deflection angle is in units of arcseconds, such deviations cannot be observed in the near future.

In Table~\ref{tab:psimg}, by using eqs.~\eqref{eqn:alpha_hat}, \eqref{eqn:lens_eq}, \eqref{eqn:impact}, and \eqref{eqn:alpha_r}, we have computed image positions, deflection angles, impact parameters, and magnifications of primary and secondary images for different values of the source angular positions. The values presented in this table hold for both a non-accelerating and slowly accelerating M87*. These quantities do not depend on the acceleration significantly. For a fixed value of the source angular position $\beta$, the impact parameter in the non-accelerating case is greater than that of the slowly accelerating case by only about 1 part in $10^{13}$. On the other hand, the (absolute value of) the image angular position is larger for the accelerating case by only about 1 part in $10^{17}$.

Therefore, it is impossible (at least   using   current and near-future observational facilities) to tell whether or not M87* is   slowly accelerating by using the image positions produced from gravitational lensing. We note that it is also impossible to tell if M87* is non-accelerating or slowly accelerating by using its shadow~\cite{Ashoorioon:2021znw}. Magnitudes of the magnification of images are greater in the slowly accelerating case by about 1 part in $10^{17}$ parts as well.

\begin{figure}[htp]
	\centering
	\includegraphics[width=0.5\textwidth]{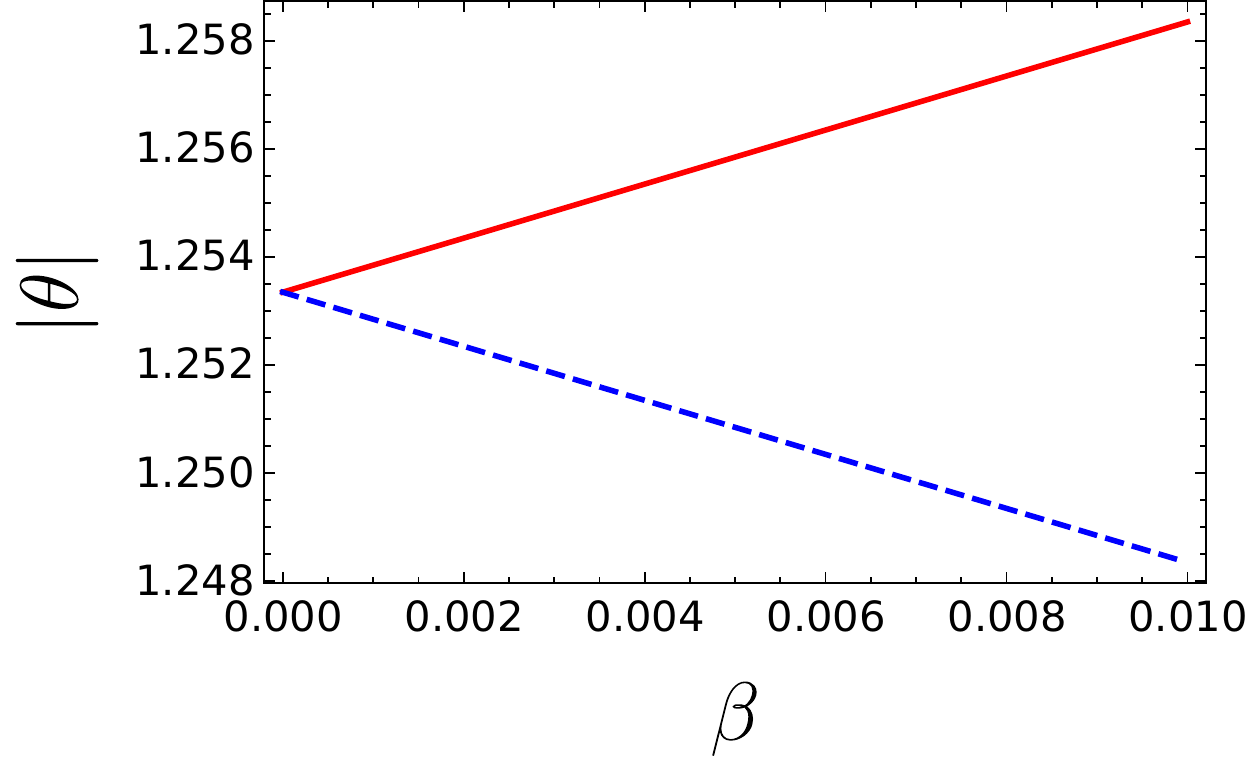}
	\caption{{\it Image position of primary and secondary images}: Angular position of primary images $\theta_{p}$ (red line) and the absolute value of angular position of secondary images $|\theta_{s}|$ (dashed blue line) as a function of source angular position $\beta$. Angles are in {\em arcseconds}. We have set $M_{{\rm M87^*}}=9.6\times 10^{12} \, {\rm m}$, $D_d=5.2\times 10^{23} \, {\rm m}$, ${\cal D}=0.5$, and $\alpha=10^{-25} {\rm m}^{-1}$.}
	\label{fig:theta_mu_vs_beta}
\end{figure}

We see from Table~\ref{tab:psimg} that the angular position of primary images increases with increasing the angular position of the source, whereas  the absolute value of the angular position of secondary images correspondingly decreases, as shown in Fig.~\ref{fig:theta_mu_vs_beta}.

\begingroup
\begin{table*}
	\caption{{\it Image positions, deflection angles, impact parameters, and magnifications of primary and secondary images due to lensing by M87*}: Angular positions $\theta$, bending angles $\hat{\alpha}$, impact parameters $b$, and magnifications $\mu$ are given for different values of angular source position $\beta$. These results are the same in the non-accelerating and slowly accelerating cases. (a) $p$ and $s$ refer to primary and secondary images, respectively. (b) All angles are in {\em arcseconds} and the impact parameters are in {\em meters}. (c) We have used $M_{{\rm M87^*}}=9.6\times 10^{12} \, {\rm m}$, $D_d=5.2\times 10^{23} \, {\rm m}$, ${\cal D}=0.5$, and $\alpha=10^{-25} {\rm m}^{-1}$.}\label{tab:psimg}
	\begin{ruledtabular}
		\begin{tabular}{l cccc cccc}
			$\beta$&$\theta_{p}$&$\hat{\alpha}_{p}$&$b_p$&$\mu_{p}$&$\theta_{s}$&$\hat{\alpha}_{s}$&$b_s$&$\mu_{s}$\\
			\hline
			$0$&$1.25334$&$2.50668$&$3.20\times 10^{18}$&$\times$&$-1.25334$&$2.50668$&$3.20\times 10^{18}$&$\times$\\
			$0.1$&$1.30436$&$2.40872$&$3.29\times 10^{18}$&$6.79185$&$-1.20436$&$2.60872$&$3.04\times 10^{18}$&$-5.78173$\\
			$0.5$&$1.52806$&$2.05611$&$3.85\times 10^{18}$&$1.82812$&$-1.02805$&$3.05611$&$2.59\times 10^{18}$&$-0.826954$\\
			$1$&$1.84943$&$1.69886$&$4.66\times 10^{18}$&$1.26694$&$-0.849483$&$3.69897$&$2.14\times 10^{18}$&$-0.267388$\\
			$2$&$2.60342$&$1.20683$&$6.56\times 10^{18}$&$1.05575$&$-0.603448$&$5.20690$&$1.51\times 10^{18}$&$-0.0567785$\\
			$3$&$3.45471$&$0.909423$&$8.71\times 10^{18}$&$1.01705$&$-0.454745$&$6.90949$&$1.14\times 10^{18}$&$-0.0176322$\\
			$4$&$4.36033$&$0.720660$&$1.10\times 10^{19}$&$1.00676$&$-0.360286$&$8.72057$&$9.08\times 10^{17}$&$-0.00687431$\\
		\end{tabular}
	\end{ruledtabular}
\end{table*}
\endgroup

\begingroup
\begin{table*}
	\caption{{\it Time delays of primary and secondary images due to lensing by M87*}: Time delays $\tau$ are given for different values of angular source position $\beta$. Instead of the time delays of secondary images $\tau_s$, we present the differential time delay $t_d=\tau_s-\tau_p$ which is of observational importance. (a) As in Table~\ref{tab:psimg}. (b) $\beta$ is in {\em arcseconds} and the (differential) time delays are in {\em seconds}. (c) As in Table~\ref{tab:psimg}. (d) Barred quantities refer to values of the case that the black hole is not accelerating and $\Delta t_d=\bar{t}_d-t_d$.}\label{tab:pst}
	\begin{ruledtabular}
		\begin{tabular}{l cc ccc}
			$\beta$&$\tau_{p}$&$t_d$&$\bar{\tau}_p$&$\bar{t}_d$&$\Delta t_d$\\
			\hline
			$0$&$3.13186970\times 10^{12}$&$0$&$1.62727362\times 10^{6}$&$0$&$0$\\
			$0.1$&$3.13186969\times 10^{12}$&$20445.4686829920$&$1.61725037\times 10^{6}$&$20445.4686829935$&$1.52795\times 10^{-9}$\\
			$0.5$&$3.13186965\times 10^{12}$&$102871.116191642$&$1.58092945\times 10^{6}$&$102871.116191651$&$8.42556\times 10^{-9}$\\
			$1$&$3.13186961\times 10^{12}$&$209681.897709557$&$1.54280388\times 10^{6}$&$209681.897709578$&$2.17115\times 10^{-8}$\\
			$2$&$3.13186955\times 10^{12}$&$448736.764965694$&$1.48443381\times 10^{6}$&$448736.764965783$&$8.89995\times 10^{-8}$\\
			$3$&$3.13186951\times 10^{12}$&$737888.423902370$&$1.44178393\times 10^{6}$&$737888.423902647$&$2.76603\times 10^{-7}$\\
			$4$&$3.13186948\times 10^{12}$&$1089214.16844197$&$1.40880628\times 10^{6}$&$1089214.168442669$&$7.01752\times 10^{-7}$\\
		\end{tabular}
	\end{ruledtabular}
\end{table*}
\endgroup

In Table~\ref{tab:pst},  we have calculated the time delay~\eqref{eqn:time_delay} of primary images, again assuming the acceleration $\alpha=10^{-25} {\rm m}^{-1}$. Quite strikingly,  even this small acceleration increases the time delay by 6 orders of magnitude, even though the deflection angles do not differ significantly. To understand this feature, recall from \eqref{eqn:tran:griffiths} that the acceleration makes important changes on the metric function only at large distances. The large distances do not have a significant contribution in the integral~\eqref{eqn:alpha_hat} due to the $1/r$ factor, and so $\alpha$ does not change the deflection angle significantly.

However, the story is different when it comes to the time delay. The integrals of Eq.~\eqref{eqn:time_delay} do not have a $1/r$ factor and  large distances consequently have a significant contribution to $\tau(b)$.
What is of observational importance is the differential time delay $t_d=\tau_s-\tau_p$ (and  $\overline{t}_d$), which we
provide  in Table~\ref{tab:pst} (instead of  explicit values of time delays of secondary images). In fact $\tau_s$ and $\tau_p$ cannot be observed, but if the source is pulsating, every phase in its period   appears in the secondary image $t_d$ seconds after it appears in the primary image.  With increasing  angular position of the source, the differential time delay $t_d$ increases.

\begin{figure}[htp]
	\centering
	\includegraphics[width=0.5\textwidth]{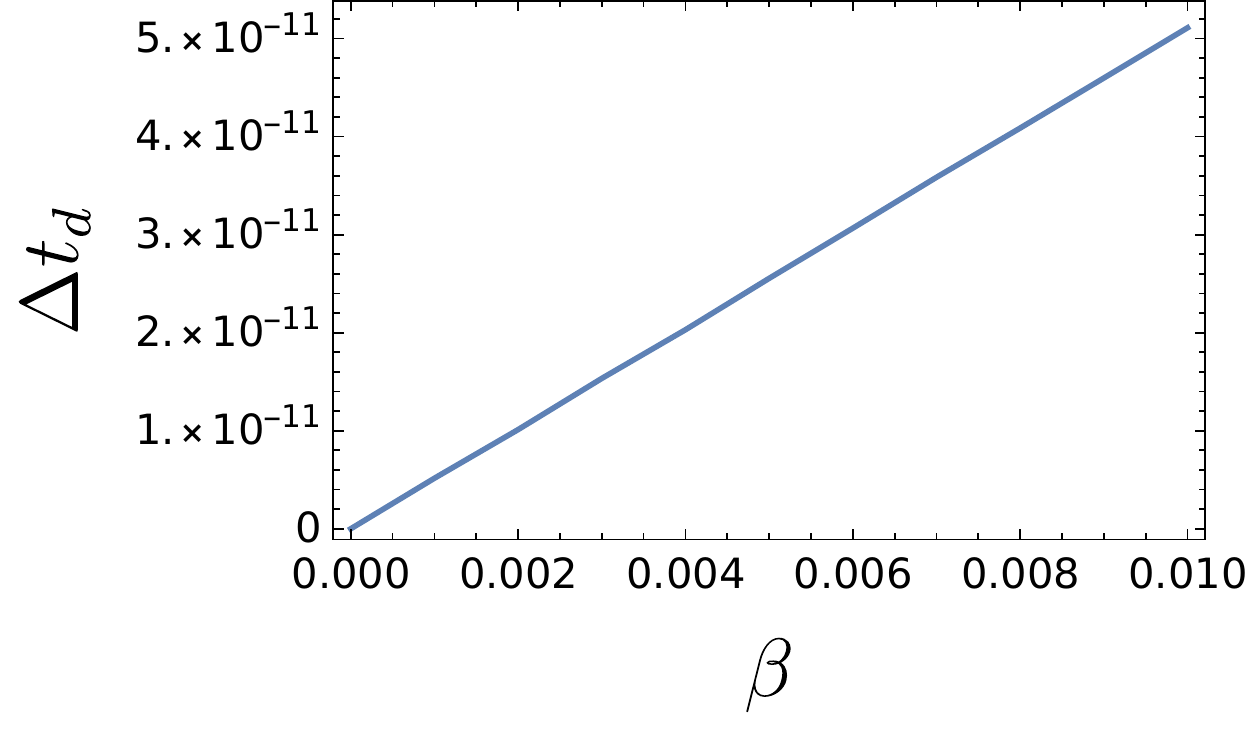}
	\caption{{\it Differential time delay}: The difference between the differential time delay in non-accelerating and slowly accelerating cases, $\Delta t_d=\bar{t}_d-t_d$. Time delays are in units of {\em seconds} and $\beta$ is in units of {\em arcseconds}. We have set $M_{{\rm M87^*}}=9.6\times 10^{12} \, {\rm m}$, $D_d=5.2\times 10^{23} \, {\rm m}$, ${\cal D}=0.5$, and $\alpha=10^{-25} {\rm m}^{-1}$.}
	\label{fig:t_d_vs_beta}
\end{figure}

Although the acceleration changes the values of $\tau_p$ (and $\tau_s$) significantly, the observable quantity $t_d$ does not deviate from its non-accelerating counterpart $\bar{t}_d$ that much. In the last column of Table~\ref{tab:pst} we have presented the difference $\Delta t_d=\bar{t}_d-t_d$, which is a positive quantity; for a fixed value of angular source position, the differential time delay of secondary and primary images is larger if the black hole is not accelerating.
 The difference increases with increasing   angular source position, as  shown in Fig.~\ref{fig:t_d_vs_beta}.

These results imply that it is indeed feasible to observe if M87* is  accelerating or not, provided sufficiently small changes $\Delta t_d$ in the value of the differential time delay can be measured. We note that Shapiro delays can be measured in binary pulsars with precisions about 10 microseconds \cite{Ng:2020uck};  precision an order of magnitude better will be required here.
Although the distance to the source (and hence ${\cal D}$) can be measured from its redshift~\cite{Schneider}, one cannot observe the angular source position $\beta$. What one sees are primary and secondary images of the source and, if the source has reliable variability, the differential time delay. Small acceleration does not change the angular positions of primary and secondary images by a feasibly observable amount,
and so the angular source position $\beta$ can be determined via Fig.~\ref{fig:theta_mu_vs_beta}.  If the differential time delay $\bar{t}_d$ corresponding to this $\beta$ matches the observed value of the differential time delay then the black hole is non-accelerating. Conversely, if this $\bar{t}_d$ does not match the observed value of the differential time delay then the black hole is slowly accelerating.

 We note that velocity of the black hole is not necessarily in the same direction as its acceleration. However, we assume the velocity to be such that the path of light from the source to the lens lie on or near the equatorial plane of the black hole. Suppose the light ray passes the black hole on its equatorial plane and starts its travel toward the observer.  The black hole is accelerating in a direction perpendicular to the equatorial plane and the geodesic equations indicate that the $\theta = \pi/2$ surface is not actually a plane with vanishing exterior curvature \cite{griffiths2006}.  
However in the small acceleration limit that we consider,  the path of the light ray deviates from the equatorial plane of the black hole by a small angle $\delta\sim \frac{\alpha D_s^2}{D_s}=\alpha D_s$ regardless of the cause of the acceleration, whether from cosmic strings \cite{hr}, magnetic fields \cite{dowker1994}, a cosmological constant \cite{mann1995}, or different combinations of these \cite{ashoorioon2021,emparan1995}.  In this limit the path of the light ray deviates from the equatorial plane of the black hole by a small angle $\delta\sim \frac{\alpha D_s^2}{D_s}=\alpha D_s$. This small deviation changes the time delays by the small fraction $\sim \alpha^2 D_s^2$. Since we have assumed that the source and the observer are all inside the acceleration horizon of the black hole and thus  $\alpha D_s \ll 1$, our result for the differential time delays will be valid up to order $\alpha^2 D_s^2$. Note also that not only $\delta\equiv \theta-\frac{\pi}{2}\ll 1$, but   $\dot{\delta} = 
\dot{\theta} \sim \theta/t \sim \alpha D_s/D_s \sim \alpha\ll D_s^{-1}$ also. In such a limit, one can show that the equation for the time delay does not change to leading order in $\delta$.

 Taking into account deviations from the equatorial plane of the black hole by an small angle $\delta$, the deflection angle  becomes
\be
\hat{\alpha}(b)=2\int_{b}^{\infty}\frac{1+\alpha m \delta}{r\sqrt{\left(\frac{r}{b}\right)^2Q_b-Q}}dr-\pi  
\ee
and we see that the time delay \eqref{eqn:time_delay} does not change to leading order in $\delta$.  Since $m \alpha$ is very small (in our example it is of order $10^{-12}$) the change in $\hat{\alpha}$ is negligible.

In gravitational lensing in the presence of an accelerating black hole, small changes accumulate over large distances from the source to the observer. We conclude that one can use gravitational lensing as a probe to measure the acceleration of a black hole that acts as a lens. We note that in the slow acceleration approximation that we are using the sign of the acceleration is not important.

 We also note that besides the primary and secondary images there is an infinite set of faint images on each side of the black hole. These are relativistic images and are produced by light rays that rotate  around the black hole before continuing their path to the observer \cite{virbhadra2000,Virbhadra:2002ju,Virbhadra:2008ws}. We postpone the study of relativistic images to a follow up publication.

\medskip

{\emph{{Acknowledgments:}}} This project has received funding /support from the European Unions Horizon 2020 research and innovation programme under the Marie Sklodowska -Curie grant agreement No 860881-HIDDeN, and was supported in part by the Natural Sciences and Engineering Council of Canada. MBJP would like to acknowledge the support of Iran Science Elites Federation and the hospitality of the University of Guilan. We are grateful to Mike Hudson and Avery Broderick for helpful correspondence.

%

\end{document}